\documentclass[journal]{IEEEtran}
\IEEEoverridecommandlockouts
\usepackage{cite}
\usepackage{amsmath,amssymb,amsfonts}
\usepackage{graphicx}
\usepackage{textcomp}
\usepackage{xcolor}
\def\BibTeX{{\rm B\kern-.05em{\sc i\kern-.025em b}\kern-.08em
    T\kern-.1667em\lower.7ex\hbox{E}\kern-.125emX}}
    
\usepackage{amsmath}
\usepackage{algorithm}
\usepackage{algorithmic}
\usepackage{hyperref}
\usepackage{booktabs}
\usepackage{tikz}
\usepackage{float}
 
\usepackage{graphicx}
\usepackage{xspace}

\usepackage{multirow}
\usepackage{caption}
\usepackage{subcaption}
\usepackage[compact]{titlesec}  
\titlespacing{\section}{2pt}{2pt}{2pt}

\usepackage{amsthm}
\newtheorem{theorem}{Theorem}
\newtheorem{lemma}{Lemma}

\newtheorem{remark}{Remark}

\newtheorem{assumption}{Assumption}
\usepackage{ upgreek }
\usepackage{soul}
\begin{document}

\title{
Escaping Barren Plateaus in Variational Quantum Algorithms Using Negative Learning Rate in Quantum Internet of Things
}

 

\author{\IEEEauthorblockN{Ratun Rahman and Dinh C. Nguyen,~\IEEEmembership{Member,~IEEE,}}
\thanks{Ratun Rahman and Dinh C. Nguyen are with the Department of Electrical and Computer Engineering, University of Alabama in Huntsville, Huntsville, AL 35899 (emails: \{rr0110,  dinh.nguyen\}@uah.edu). }

}
\maketitle
\pagenumbering{gobble} 



\begin{abstract}
Variational Quantum Algorithms (VQAs) are becoming the primary computational primitive for next-generation quantum computers, particularly those embedded as resource-constrained accelerators in the emerging Quantum Internet of Things (QIoT).  However, under such device-constrained execution conditions, the scalability of learning is severely limited by barren plateaus, where gradients collapse to zero and training stalls.  This poses a practical challenge to delivering VQA-enabled intelligence on QIoT endpoints, which often have few qubits, constrained shot budgets, and strict latency requirements. In this paper, we present a novel approach for escaping barren plateaus by including negative learning rates into the optimization process in QIoT devices.  Our method introduces controlled instability into model training by switching between positive and negative learning phases, allowing recovery of significant gradients and exploring flatter areas in the loss landscape. We theoretically evaluate the effect of negative learning on gradient variance and propose conditions under which it helps escape from barren zones.  The experimental findings on typical VQA benchmarks show consistent improvements in both convergence and simulation results over traditional optimizers.  By escaping barren plateaus, our approach leads to a novel pathway for robust optimization in quantum-classical hybrid models.
\end{abstract}

\maketitle

\begin{IEEEkeywords}
Barren Plateaus, negative learning rate, variational quantum algorithms
\end{IEEEkeywords}

\section{Introduction} \label{Sec:Introduction}
\textcolor{black}{In the emerging \textbf{Quantum Internet of Things (QIoT)}, quantum processing units (QPUs) and portable quantum accelerators are embedded into edge devices to perform real-time sensing, local inference, and lightweight optimization under stringent resource constraints \cite{adil2025quantum, panahi2025energy}.  Unlike large laboratory quantum computers, these QIoT endpoints often have a modest number of qubits, limited shot budgets, restricted memory, significant readout noise, and severe latency constraints.  As a result, learning must be accomplished quickly on-device, without the usage of deep expressive circuits or repetitive complete re-initialization cycles, which would significantly increase execution time and power consumption.  QIoT highlights the need for \emph{optimizer-level} methods that enable quantum learning to be trainable even with limited depth and sampling overhead.
Variational quantum algorithms (VQAs) \cite{cerezo2021variational} provide the most viable computational foundation for near-term QIoT systems.  A QIoT device runs a parameterized quantum circuit (PQC) locally, while a classical optimizer changes parameters based on observed statistics. This allows hybrid quantum-classical routines to perform tasks like anomaly detection, parameter estimation, and adaptive sensing on low-resource hardware.  Importantly, this workflow takes use of the fact that only scalar measurement information or model parameters must be shared externally, rather than entire quantum states, reducing quantum communication cost and making VQA-style processing compatible with edge-level QIoT designs.}

Despite their potential, VQAs suffer from obstacles, particularly the phenomenon of barren plateaus, in which the gradients of the optimization landscape disappear rapidly as the number of qubits grows \cite{mcclean2018barren, cerezo2021cost}. In a barren plateau, the cost function becomes almost flat in large areas of the parameter space, making it exceedingly difficult for gradient-based optimization methods to determine a meaningful direction for parameter updates \cite{arrasmith2021effect}. As a result, the training process stalls out and may require exponentially many observations to properly predict gradients, making optimization essentially impossible for large-scale systems \cite{holmes2022connecting}. This problem arises from the high-dimensional geometry of quantum states and the concentration of measurement phenomena encountered in random quantum circuits \cite{pesah2021absence}. Basic QML approaches, such as standard gradient descent or heuristic optimization, cannot overcome barren plateaus because they rely on the existence of relevant gradient data \cite{wang2021noise}. Without structural modifications to the algorithm or parameterized circuit design, these techniques have inherent limitations in their ability to escape barren zones and perform scalable quantum learning \cite{grant2019initialization, volkoff2021large}. To overcome the limitations imposed by barren plateaus in VQA \textcolor{black}{in QIoT devices}, we present a novel approach that employs negative learning rates during VQA training.  We can summarize the contribution as follows.


\begin{figure*}
    \centering
    \begin{subfigure}[t]{0.40\textwidth}
        \centering
        \includegraphics[width=\linewidth]{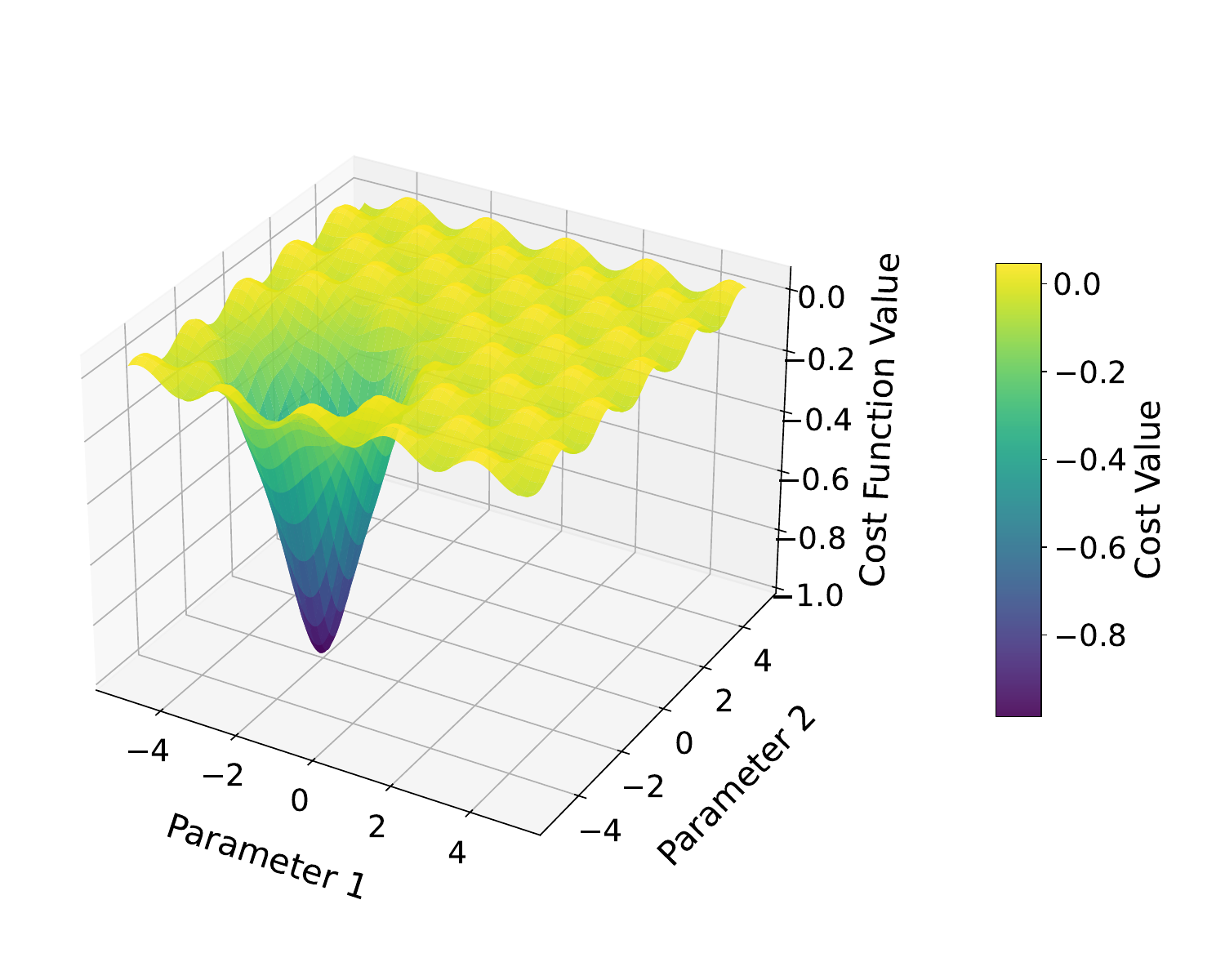}
        \caption{\footnotesize  3D surface of the barren plateau landscape. A shallow cave at $(-2, -2)$ creates a flat optimization challenge.}
        \label{fig:sub-barren3d}
    \end{subfigure}
    \hfill
    \begin{subfigure}[t]{0.40\textwidth}
        \centering
        \includegraphics[width=\linewidth]{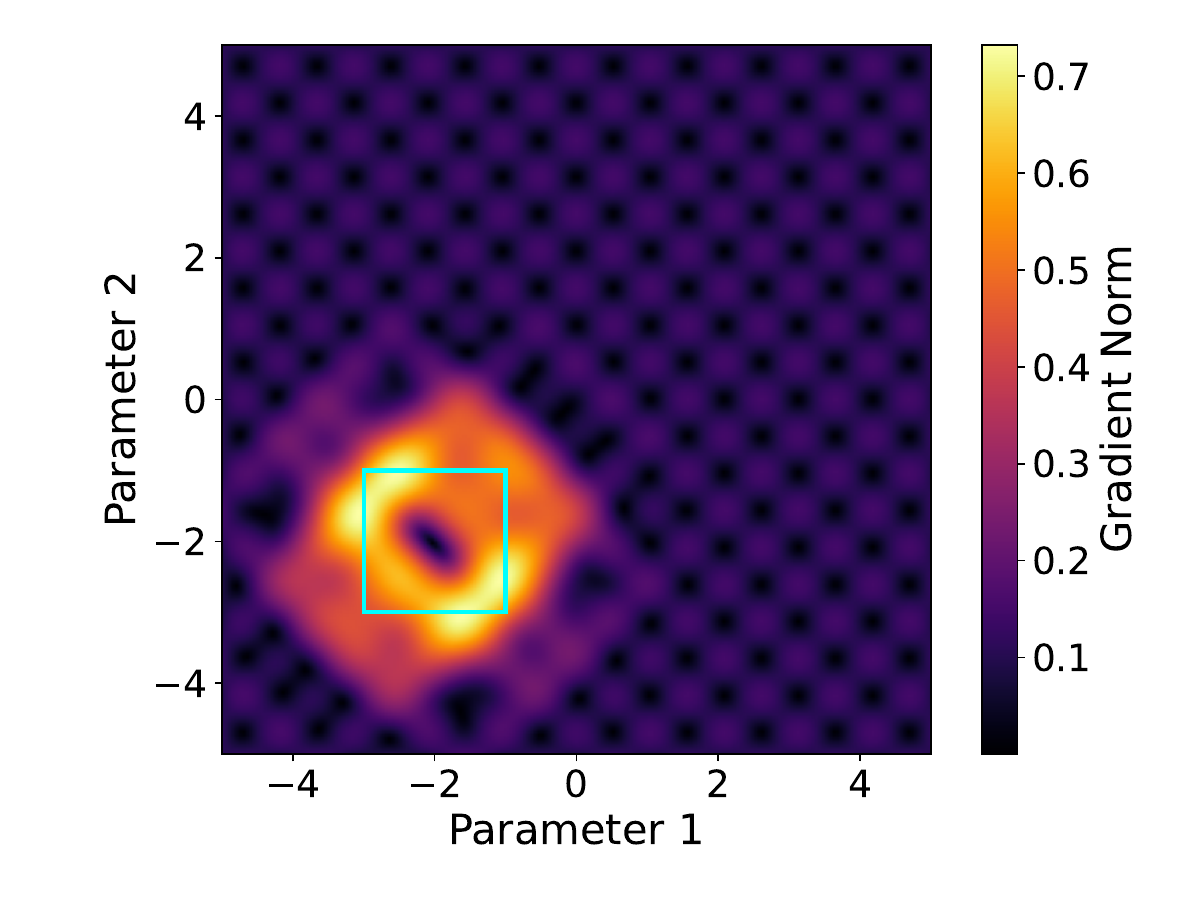}
        \caption{\footnotesize  Gradient magnitude heatmap. The cyan box highlights the barren plateau near $(-2, -2)$.}
        \label{fig:sub-barren2d}
    \end{subfigure}
    \caption{\footnotesize  Visualization of a synthetic barren plateau. (a) shows the 3D cost surface with a cave structure centered at $(-2, -2)$, while (b) illustrates the corresponding gradient vanishing behavior.}
    \label{fig:combined-barren}
    \vspace{-1em}
\end{figure*}

\begin{itemize}
    \item We propose a novel approach of negative learning rates in the context of VQA training \textcolor{black}{for QIoT devices}, laying the theoretical groundwork for their effectiveness in overcoming barren plateaus by implementing an effective training method that alternates between positive and negative learning phases to promote gradient amplification and improve landscape exploration.
    \item We investigate the behavior of negative learning phases and explain when and why negative learning might successfully decrease barren plateau problems.
    \item We perform extensive experiments in VQAs and show that our approach consistently and reliably improves the convergence and performance of the model compared to traditional optimization approaches, reducing the classification loss by up to \textbf{8.2\%} across both synthetic and publicly available datasets.
\end{itemize}

\section{Related works}
\textcolor{black}{The phenomenon of barren plateaus is widely acknowledged as a critical barrier to scalable VQA training. McClean et al.~\cite{mcclean2018barren} demonstrated that in deep unstructured parameterized quantum circuits, the gradients of the cost function vanish exponentially with the number of qubits. This makes gradient–based optimization ineffective as system size grows, which is particularly restrictive for low–resource QIoT devices where circuits cannot be made arbitrarily deep and repeated re–initialization cycles are impractical. Beyond optimization–level studies, recent work has explored how quantum technologies integrate with IoT architectures, leading to the emergence of QIoT. Adil et al.~\cite{adil2025quantum} examined the effect of quantum computing on healthcare IoT, showing how quantum–enhanced communication and learning can strengthen trust and privacy. Panahi~\cite{panahi2025energy} introduced an energy–efficient and decoherence–aware entanglement generation architecture for resource–limited QIoT endpoints running VQA–based workloads. These results collectively highlight the growing importance of lightweight, trainable quantum learning pipelines in IoT settings, motivating our work toward optimizer–level stabilization of variational models for QIoT devices.
}

Several subsequent studies have helped to learn the reasons for barren plateaus.  By differentiating between global and local cost functions, Cerezo et al.\cite{cerezo2021cost} showed that the selection of the cost function has a substantial impact on the formation of barren plateaus.  Global cost functions, which need measurements throughout the whole system, are more likely to produce barren plateaus, but local cost functions, which focus on particular subsystems, can reduce gradient vanishing.  According to Grant et al.\cite{grant2019initialization}, initialization techniques are also important; although well-planned layerwise or identity-preserving initializations might enhance trainability, random parameter initialization often results in barren plateaus.
The entanglement architecture of the parameterized circuits also determines the appearance of barren plateaus.  Marrero et al. \cite{ortiz2021entanglement} showed that excessive entanglement causes the concentration of measurement phenomena, increasing the disappearance of the gradient.  Similarly, Pesah et al.\cite{pesah2021absence} investigated how organized, shallow circuits could prevent barren plateaus by restricting entanglement development. Noise and hardware imperfections can also result in barren plateaus.  Wang et al.~\cite{wang2021noise} investigated how noise contributes to the problem of vanishing gradient, resulting in "noise-induced barren plateaus."  Their findings indicate that even professionally planned ansätze can suffer from barren plateaus in realistic noise situations \cite{rahman2025sporadic}.

Several mitigation strategies have been proposed to address the problems of barren plateaus. Designing shallow, problem-inspired approaches~\cite{grimsley2023adaptive}, using hardware-efficient circuits~\cite{sim2019expressibility}, and the implementation of local cost functions~\cite{cerezo2021cost} have shown promising results.  Adaptive techniques, including dynamic circuit growth~\cite{volkoff2021large} and symmetry-informed pruning~\cite{arrasmith2022equivalence}, have been developed. Skolik et al. ~\cite{skolik2021layerwise} proposed a layer-wise learning strategy for quantum neural networks, gradually increasing the depth of the circuit during training to maintain training capacity and prevent gradient vanishing in deeper networks. 
\textcolor{black}{More recent research has looked into various approaches to addressing barren plateaus. 
 Zhang et al.~\cite{zhang2022escaping} suggested Gaussian-based initializations to decrease gradient decay in deeper circuits, whereas Yao and Hasegawa~\cite{yao2025avoiding} studied the role of entanglement structure in avoiding plateaus entirely. 
 Liu et al. ~\cite{liu2025stochastic} found that well-specific perturbations can enhance optimization-level techniques, including negative learning phases.}
\textcolor{black}{In addition, a few studies have investigated random restarts / random re-initialization procedures as a means of avoiding flat sections in VQA \cite{rahman2025towards}.  The main concept is to continually re-sample parameters and re-run optimization until a non-plateau zone is located. This can assist in shallow instances, but it does not preserve curvature information throughout training \cite{arrasmith2021effect}. In contrast, our NLR method adds gradient-aligned reversals within the same local area instead of discarding accumulated structure. This allows for more systematic escape dynamics than re-initialization-based sampling.}

Despite these developments, a widely applicable optimization-based strategy for systematically escaping barren plateaus remains elusive.  Most current mitigation strategies focus on architectural changes or initialization schemes rather than substantially affecting optimization dynamics. This gap encourages the investigation of alternative optimization techniques, such as the integration of negative learning phases, in order to improve trainability without needing significant modifications to the circuit or problem structure.

\section{Methods}
\subsection{Quantum Machine Learning}
Each client constructs a local quantum model using a VQA applied to an encoded quantum state. A VQA consists of $L$ layers, each comprising parameterized single-qubit rotation gates (e.g., $R_x$, $R_y$, $R_z$) that act on individual qubits, together with entangling gates (e.g., CNOT) that generate correlations between qubits. The overall unitary transformation at round $t$ is written as
\begin{equation}
    U(\theta_{t}) = \prod_{\ell=1}^L U_\ell(\theta_{t,\ell}), 
    \label{eq:VQA_unitary}
\end{equation}
where $\theta_{t}$ is the parameter vector at round $t$, and $U_\ell(\theta_{t,\ell})$ denotes the unitary operation in layer $\ell$, parameterized by $\theta_{t,\ell}$.  
If $G_\ell$ is the Hermitian generator (e.g., a Pauli operator) associated with a rotation gate, then the unitary operator for layer $\ell$ is
\begin{equation}
    U_\ell(\theta_{t,\ell}) = \exp\left(-i \tfrac{\theta_{t,\ell}}{2} G_\ell\right).
\end{equation}

After applying the VQA, the output quantum state is obtained as
\begin{equation}
    |\psi_{\text{out}}(\theta, \theta_{t})\rangle 
    = U(\theta_{t})\, |\psi_{\text{enc}}(\theta)\rangle,
    \label{eq:VQA_output_state}
\end{equation}
where $|\psi_{\text{enc}}(\theta)\rangle$ denotes the encoded input state. To produce a prediction, the state is measured with respect to a Hermitian observable $O$ (e.g., Pauli-$Z$), yielding the expectation value
\begin{equation}
    f(\theta, \theta_{t}) 
    = \langle \psi_{\text{out}}(\theta, \theta_{t}) | O | \psi_{\text{out}}(\theta, \theta_{t}) \rangle.
    \label{eq:measurement_expectation}
\end{equation}

In practice, measurements are repeated $M$ times (shots) to obtain an empirical estimate
\begin{equation}
    \hat{f}_{t}(\theta, \theta_{t}) 
    = \frac{1}{M} \sum_{j=1}^{M} H_j,
\end{equation}
where $H_j$ denotes the observed measurement outcome of the $j$-th shot. For a given data point $(x,y)$, the prediction $\hat{f}_{t}(\theta, \theta_{t})$ is compared against the label $y$ through a loss function $\ell(y,\hat{f}_{t}(\theta,\theta_{t}))$. For each local epoch $t$, a mini-batch $d_t \subseteq D_t$ of size $|d_t|$ is sampled from the client dataset, and the mini-batch loss is defined as
\begin{equation}
    L(\theta_{t}) 
    = \frac{1}{|d_t|} \sum_{(x,y)\in d_t} 
    \ell\big(y, \hat{f}_{t}(\theta,\theta_{t})\big).
    \label{eq:batch_loss}
\end{equation}

Gradients of the expectation value with respect to circuit parameters are computed using the parameter-shift rule. For the $d$-th parameter,
\begin{equation}
    \frac{\partial f(\theta, \theta_{t})}{\partial \theta_{t,d}} 
    = \tfrac{1}{2}\left[ 
        f(\theta, \theta_{t} + \tfrac{\pi}{2} e_d) 
        - f(\theta, \theta_{t} - \tfrac{\pi}{2} e_d) 
    \right],
    \label{eq:parameter_shift}
\end{equation}
where $e_d$ is the unit vector along the $d$-th coordinate of the parameter space.

Throughout this paper, we refer to the client-level mini-batch loss in Eq.~\eqref{eq:batch_loss} as the \emph{local loss function} $L(\theta_t)$, while the term \emph{global cost function} $C(\theta_t)$ denotes the aggregated objective across all participating clients.

\subsection{Negative Learning Rate Training}
Classical optimization of VQA typically relies on gradient descent, where parameters are updated in the opposite direction of the gradient. This assumes that the gradient consistently points toward lower-cost regions. However, in barren plateau regimes, gradients vanish exponentially with system size~\cite{mcclean2018barren}, leaving the optimizer with almost no information about how to adjust parameters. As a result, standard gradient descent often stalls.
To address this issue, we introduce \emph{negative learning-rate (NLR) training}, an optimization strategy that incorporates controlled gradient reversals to encourage exploration. Instead of exclusively following the negative gradient, NLR permits occasional updates in the \emph{positive} gradient direction, thereby implementing a form of cost-conditioned ascent. The intuition is that when gradients are extremely small, reversing the update direction perturbs the parameter trajectory, enabling the model to escape flat regions and reach neighborhoods with more informative gradients.

The procedure operates as follows. At each step, a tentative gradient descent update is performed and its cost is evaluated:
\begin{equation}
    \theta_t' = \theta_t - \eta \nabla_\theta C(\theta_t), \quad 
    \Delta C = C(\theta_t') - C(\theta_t).
\end{equation}
\begin{itemize}
    \item If $\Delta C \le 0$ (the cost decreases or remains the same), the update is accepted: $\theta_{t+1} = \theta_t'$.
    \item If $\Delta C > 0$ (the cost increases), the update is reversed with a \emph{negative learning rate}:
    \begin{equation}
        \theta_{t+1} = \theta_t + \eta' \nabla_\theta C(\theta_t),
    \end{equation}
    where $\eta' > 0$ controls the magnitude of the ascent step.
\end{itemize}

Thus, NLR training alternates between standard descent steps and occasional controlled ascent steps triggered by failed descent attempts. This mechanism transforms the optimization trajectory into a stochastic exploration process that performs a random walk in barren plateau regions. 

\textcolor{black}{
\textbf{Guideline for Selecting the Negative Learning Rate.}
Choosing a suitable negative learning rate $\eta'$ is crucial for achieving stable yet exploratory optimization.
 $\eta$ defines the conventional descent scale, but $\eta'$ specifies how strongly the model reacts as the cost increases.
 A useful heuristic is to set $\eta' = \kappa \eta$, while $\kappa \in [1.5, 3.0]$ depends on the circuit depth $L$, Hamiltonian variance $\sigma_H^2$, and hardware noise rate $\nu$ as
\begin{equation}
    \eta' = \eta \left(1 + \log(1 + \nu L)\right) \sqrt{\frac{\sigma_H^2}{L}}.
\end{equation}
Smaller numbers ($\eta' \approx 1.5\eta$) are appropriate for shallow, low-noise circuits, but bigger ratios ($\eta' \approx 3\eta$) are advantageous for deeper or noisier settings.
In practice, an initial $\eta' = 2\eta$ gives a reliable starting point that can be changed based on loss oscillations or divergence behavior.
This adaptive algorithm provides a reproducible mechanism for tuning $\eta'$ based on the Hamiltonian, ansatz structure, circuit depth, and noise profile.
}
\subsection{Algorithm}
\begin{algorithm}[t]
\caption{\footnotesize  Negative Learning-Rate Training for Variational Quantum Circuits}
\label{alg:nl-vqa}
\small
\begin{algorithmic}[1]
\REQUIRE Initial parameters $\theta_0$, learning rates $\eta>0$, $\eta'>0$, total steps $T$, mini-batch size $B$
\ENSURE Final parameters $\theta_T$
\STATE Initialize cost $C(\theta_0)$ \label{line:init_loss}
\FOR{$t = 0, \ldots, T-1$}
    \STATE Sample mini-batch $d_t \subseteq D$ of size $B$ \label{line:sample_batch}
    \STATE Compute mini-batch loss $L_t(\theta_t)$ \label{line:compute_loss}
    \STATE Estimate gradient $g_t \gets \nabla_\theta L_t(\theta_t)$ (e.g., by parameter-shift rule) \label{line:compute_gradient}
    \STATE Tentative descent step: $\theta_{\text{temp}} \gets \theta_t - \eta \cdot g_t$ \label{line:tentative_update}
    \STATE Evaluate mini-batch loss: $L_{\text{temp}} \gets L_t(\theta_{\text{temp}})$ \label{line:compute_temp_loss}
    \IF{$L_{\text{temp}} > L_t(\theta_t)$} \label{line:check_loss}
        \STATE Negative learning step: $\theta_{t+1} \gets \theta_t + \eta' \cdot g_t$ \label{line:reverse_update}
    \ELSE
        \STATE Accept descent step: $\theta_{t+1} \gets \theta_{\text{temp}}$ \label{line:accept_update}
    \ENDIF
    \STATE Update cost estimate: $C(\theta_{t+1}) \gets$ aggregate loss evaluated at $\theta_{t+1}$ \label{line:update_loss}
\ENDFOR
\end{algorithmic}
\end{algorithm}

To avoid barren plateaus, we investigate training a single quantum variational model using mini-batch stochastic gradient optimization and loss-driven negative learning rate steps. Algorithm~\ref{alg:nl-vqa} explains the suggested negative learning rate training approach.
Initially (Line~\ref{line:init_loss}), the cost function is assessed at the initial parameters $\theta_0$. Each training step involves sampling a mini-batch from the dataset (Line~\ref{line:sample_batch}) and computing the related mini-batch loss.
The gradient for the mini-batch is calculated (Line~\ref{line:compute_gradient}), followed by a preliminary standard gradient descent update (Line~\ref{line:tentative_update}). The loss after the tentative update is measured (Line~\ref{line:compute_temp_loss}). If the tentative update produces a worse loss than the previous model (Line~\ref{line:check_loss}), a negative learning phase is initiated, and the update is reversed in the gradient direction (Line~\ref{line:reverse_update}). Otherwise, tentative updates are accepted (Line~\ref{line:accept_update}). The revised cost function of the model is then calculated for future comparison (Line~\ref{line:update_loss}). This dynamic alternating between positive and negative learning stages allows the model to go beyond barren plateaus and seek more effective optimization routes.

\section{Theoretical Analysis}
VQAs suffer severe training challenges due to \emph{barren plateaus}, where gradient magnitudes drop exponentially with the number of qubits, leading to near-zero updates in gradient-based optimization \cite{mcclean2018barren}.  The strong expressivity of deep circuits results in the cost function's variance decaying as $O(2^{-d})$ for $d$ dimensions of qubit \cite{cerezo2021cost}.  We present a comprehensive theoretical explanation of how \emph{NLR} training mitigates these challenges by producing loss-conditioned non-monotone steps, increasing exploration in flat landscapes, and lowering prediction time to escape barren plateaus.  Unlike techniques that combine two positive step sizes, NLR uses sign flips to sample opposite directions, reducing noise-induced errors and improving diffusion.


\subsection{Assumptions}
We use the following requirements, which are conventional in non-convex optimization and suited to quantum situations with noisy gradients

\begin{assumption}[Smoothness]
\label{ass:smoothness}
The cost function $C: \mathbb{R}^d \to \mathbb{R}$ is twice continuously differentiable, with $L$-Lipschitz gradients (i.e., $\|\nabla C(\theta) - \nabla C(\theta')\| \leq L \|\theta - \theta'\|$) and bounded Hessian $\|\nabla^2 C(\theta)\| \leq L$ for every $\theta \in \mathbb{R}^d$.
\end{assumption}

\begin{assumption}[Barren Plateau Region]
\label{ass:plateau}
A barren plateau area $\mathcal{B} = \{\theta: \|\nabla C(\theta)\| \leq \varepsilon, \|\nabla^2 C(\theta)\| \leq L\}$ exists, where $\varepsilon = O(\exp(-\alpha d))$ for dimentions $d$ and constant $\alpha > 0$.
\end{assumption}

\begin{assumption}[Stochastic Gradients]
\label{ass:gradients}
The observed gradient is $g_t = \nabla C(\theta_t) + \xi_t$, where $\xi_t$ is noise with $\mathbb{E}[\xi_t|\theta_t] = 0$ and covariance $\mathrm{Cov}(\xi_t|\theta_t) = \Sigma_t \preceq \sigma^2 I$, with $\sigma^2 = O(\exp(-\alpha d))$ representing finite-shot measurement noise in VQAs.
\end{assumption}

\begin{assumption}[Small Step Sizes]
\label{ass:step-sizes}
For learning rates $\eta, \eta' > 0$, $\max(\eta, \eta') \leq \eta_{\max}$, where $\eta_{\max}$ is small enough that Taylor remainders are $o(\eta^2 \|g_t\|^2)$.
\end{assumption}

\begin{assumption}[Acceptance Rule]
\label{ass:acceptance}
At iteration $t$, calculate a tentative descent $\theta_t' = \theta_t - \eta g_t$.  If $C(\theta_t') \leq C(\theta_t)$, define $\theta_{t+1} = \theta_t'$.  Otherwise, let $\theta_{t+1} = \theta_t + \eta' g_t$.  For simplicity, it is assumed that cost assessments are accurate; noisy evaluations raise the likelihood of a violation.
\end{assumption}

These assumptions capture the flat, noisy landscapes of barren plateaus while maintaining analytical tractability.

\subsection{Technical Lemmas}
We first describe the conditions under which the attempted descent fails, connect them to local curvature and noise.

\begin{lemma}[Acceptance Test via Curvature]
\label{lem:accept}
For $\theta_t \in \mathcal{B}$ and a small $\eta$, the change of cost is:
\begin{equation}
C(\theta_t - \eta g_t) - C(\theta_t) = -\eta g_t^\top \nabla C(\theta_t) + \frac{\eta^2}{2} g_t^\top H_t g_t + O(\eta^3 \|g_t\|^3),
\end{equation}
where $H_t=\nabla^2 C(\theta_t)$.  The violation event $\mathcal{E}_t = \{C(\theta_t - \eta g_t) > C(\theta_t)\}$ happens approximately when
\begin{equation}
\frac{\eta}{2} g_t^\top H_t g_t > g_t^\top \nabla C(\theta_t).
\end{equation}
\textcolor{black}{In plateau areas, where $\|\nabla C(\theta_t)\| \leq \varepsilon \ll \sigma$, the first-order gradient signal is severely weak.   The apparent cost variation may be impacted by both curvature and stochastic noise, depending on the local size of $\|H_t\|$.  If the curvature is considerable or combined with noise fluctuations, these effects may dominate the step outcome, resulting in a failure descent.  
 When both the gradient and the Hessian are minimal, updates are practically neutral, with little change in cost.}
\end{lemma}

\textit{Proof Sketch.} Use the noisy gradient from Assumption~\ref{ass:gradients} to apply the Taylor expansion under Assumption~\ref{ass:smoothness}.  
\textcolor{black}{The violation occurs when the quadratic component exceeds the weak linear decline term, which is affected by noise in the positive curvature directions. This does not imply significant curvature everywhere, but rather emphasizes that modest curvature or noise may momentarily dominate in flat areas.}  
Full proof in Appendix~\ref{app:proofs}.

\textcolor{black}{This updated explanation shows that tentative descents may fail not because curvature is inherently enormous, but because very small gradients mixed with curvature or noise unpredictability can obscure descent directions, a phenomenon common in barren plateaus.}

\begin{lemma}[Backtracking Cap under Violation]
\label{lem:cap}
Assuming $\mathcal{E}_t$ occurs,  $C(\theta_t - \tilde\eta_t g_t) \leq C(\theta_t) - c \tilde\eta_t \|g_t\|^2$ (e.g., Armijo condition, $c > 0$) is a descent-only backtracking rule that requires:
\begin{equation}
\tilde\eta_t \leq \frac{2(1-c) g_t^\top \nabla C(\theta_t)}{g_t^\top H_t g_t} + O(\tilde\eta_t^2 \|g_t\|).
\end{equation}
In plateaus, $\tilde\eta_t \lesssim O(\varepsilon / (\sigma^2 d)) \ll \eta$ if curvature is non-negligible.
\end{lemma}

\textit{Proof Sketch.} Derive the Armijo inequality by solving for the maximum $\tilde\eta_t$ using the Taylor expansion from Lemma~\ref{lem:accept}.  The constrained curvature and small gradients enable the bound to diminish in $\mathcal{B}$. Full proof in Appendix~\ref{app:proofs}.

This lemma emphasizes that in violation scenarios, backtracking drastically reduces steps, which inhibits exploration in noisy plateaus.

\subsection{Diffusion in Barren Plateaus}
Next, we contrast the diffusive behavior of NLR with a backtracking technique that relies solely on descent.

\begin{theorem}[Diffusion Ordering in Barren Plateaus]
\label{thm:diffusion}
Suppose that $p_t = \mathbb{P}(\mathcal{E}_t|\theta_t)$.  For $\mathcal{B}$, the NLR update is expressed as
\begin{equation}
\Delta\theta_t^{\mathrm{NLR}} = \begin{cases}
-\eta g_t, & \text{if } \neg \mathcal{E}_t, \\
+\eta' g_t, & \text{if } \mathcal{E}_t,
\end{cases}
\end{equation}
while the backtracking update is formulated as
\begin{equation}
\Delta\theta_t^{\mathrm{BT}} = \begin{cases}
-\eta g_t, & \text{if } \neg \mathcal{E}_t, \\
-\tilde\eta_t g_t, & \text{if } \mathcal{E}_t.
\end{cases}
\end{equation}
The diffusion coefficients, assuming isotropic noise ($\Sigma_t = \sigma^2 I / d$), are described as 
\begin{equation}
\begin{aligned}
    D^{\mathrm{NLR}} &= \left[ (1-p_t) \eta^2 + p_t (\eta')^2 \right] \sigma^2, \\
D^{\mathrm{BT}} &= \left[ (1-p_t) \eta^2 + p_t \mathbb{E}[\tilde\eta_t^2 | \mathcal{E}_t] \right] \sigma^2.
\end{aligned}
\end{equation}
If $\eta' > \mathbb{E}[\tilde\eta_t | \mathcal{E}_t]$, then $D^{\mathrm{NLR}} > D^{\mathrm{BT}}$.
\end{theorem}

\textit{Proof Sketch.} The calculation is $\mathbb{E}[\|\Delta \theta_t\|^2]$ for both approaches, estimating conditional expectations in plateaus with almost isotropic $g_t$.  While NLR's constant $\eta'$ preserves greater variance, backtracking's reduced $\tilde\eta_t$ (from Lemma~\ref{lem:cap}) decreases $D^{\mathrm{BT}}$. Full proof in Appendix~\ref{app:proofs}.

This theorem demonstrates that, in contrast to the conservative shrinking of backtracking, NLR's sign flip permits longer steps in violation directions, improving exploration.

\begin{remark}[Exit-Time Scaling]
\color{black}
\label{lem:exit}
In barren plateau areas, parameter updates with small step sizes $\eta_{\max}$ behave like a discrete random walk accompanied by stochastic gradient noise. 
 As $\eta_{\max} \!\to\! 0$, this trajectory is approximated by a continuous diffusion process with effective coefficient $D$. 
 With this approximation, the predicted time to depart a plateau region of radius $R$ grows approximately.
\(
    \mathbb{E}[\tau_{\text{exit}}] \approx \frac{R^2}{2D},\)s
corresponds to the average first-passage time of Brownian motion in $d$-dimensional space~\cite{schuss2009theory}. 
This interpretation, while not a precise limit theorem, provides an intuitive notion that a higher diffusion coefficient—achieved through negative learning phases—leads to faster escape from flat loss landscapes.
\end{remark}



\subsection{Post-Escape Behavior}
NLR operates similarly to regular SGD after gradients are no longer on the plateau, where they become informative.

\begin{theorem}[Post-Escape Stability]
\label{thm:stability}
Assume $\mathcal{G}_\varepsilon = \{\theta: \|\nabla C(\theta)\| \geq \varepsilon\}$.  For $\theta_t \in \mathcal{G}_\varepsilon$, the violation probability $p_t \to0$.  The approach approximates stochastic gradient descent using infrequent disturbances.  Using a Robbins--Monro schedule ($\sum_t \eta_t = \infty$, $\sum_t \eta_t^2 < \infty$), the iterates nearly certainly reach stationary positions $\{\theta: \nabla C(\theta) = 0\}$.  Given constant step sizes, $\mathbb{E}[\|\nabla C(\theta_t)\|^2] \leq O(\eta \sigma^2 + 1/t)$.
\end{theorem}

\textit{Proof Sketch.} In informative regions, linear descent dominates (Lemma~\ref{lem:accept}), which reduces $p_t$.  Convergence follows traditional SGD results with bounded variance, and rare ascents do not interrupt the process. Full proof in Appendix~\ref{app:proofs}.

\begin{remark}
NLR's sign flip provides directional diversity, reducing noise-induced inaccuracies in barren plateaus, as opposed to two positive rates that remain in the same direction (scaled).  This is consistent with observations that stochastic perturbations aid escape in quantum environments \cite{liu2025stochastic}.  However, for deep PQCs, exponential scaling remains \cite{larocca2025barren}, restricting NLR to shallow circuits unless supplemented with additional mitigations, such as structured initializations \cite{grant2019initialization}.
\end{remark}

\section{Experiments}
\subsection{Quantum Model Architecture}
The quantum model used in our research is a PQC with six qubits.  Amplitude encoding maps classical input vectors to quantum states.  The PQC has $L=5$ levels, with each layer using a series of trainable single-qubit rotation gates ($R_x$, $R_y$, and $R_z$) followed by a set of CNOT gates to entangle neighboring qubits.  Each rotation gate is defined by a learnable angle.  The entanglement structure is linear, with each qubit connected to its nearest neighbor to optimize the circuit depth efficiency. Following unitary evolution, the final quantum state is measured using a Hermitian observable, especially the Pauli-Z operator, and the expectation value is utilized as the model's prediction output.  The entire design achieves an appropriate balance between expressibility and trainability while limiting circuit depth to prevent excessive noise amplification. 

\textcolor{black}{\textbf{Experimental Environment.} 
Training employs mini-batch stochastic gradient descent with the proposed negative learning rate scheme. 
 Each iteration samples a batch of size $B=32$, conducts a forward update with $\eta=0.01$, then reverses direction with $\eta'=0.02$ when the loss grows. 
 Measurements employ $M=1000$ shots, and training lasts $T=500$ steps with dynamic switching between positive and negative phases. 
 The model parameters are evenly initialized from $[-\pi,\pi]$; classical layers adopt Xavier initialization; and noise and switching thresholds are controlled by constants $\sigma=0.05$ and $\tau=0.1$.
 The simulations are run on an NVIDIA RTX 4090 GPU with 64GB RAM running Ubuntu 22.04.}

\subsection{Dataset}
In our primary experiments, we create a customized synthetic dataset with properties important to investigating barren plateaus in variational quantum optimization.  Two multivariate Gaussian distributions are constructed in $\mathbb{R}^d$, each representing a different class.  Samples are taken at random, normalized to the unit norm, and then mapped to quantum states using amplitude encoding.  This encoding guarantees that the data fully leverage the Hilbert space structure, resulting in complicated cost landscapes.  The categorization labels are assigned according to the original Gaussian distribution.  The created dataset is partitioned into a training set (80\%) and a test set (20\%), with no data augmentation techniques performed. This is an ideal condition for barren plateaus because: i)
High dimensionality and non-linearity are introduced by amplitude encoding.
ii) The Hilbert space is filled with normalized vectors sampled at random.
iii) Optimization becomes difficult, resembling the behavior of plateaus.
This controlled synthetic setting enables us to carefully study the optimization behavior at different levels of barren plateau severity.

In addition to synthetic data, we have also used some publicly available datasets.  The \textit{Quantum Data Set (QDataSet)}~\cite{perrier2022qdataset} labels synthetic quantum states for supervised learning tasks and has been used to assess variational algorithms.  The \textit{MNIST} dataset~\cite{lecun1998gradient}, initially classical, has been extensively utilized in quantum investigations after pre-processing (e.g., by lowering image dimensionality and encoding quantum states).  The \textit{Fashion-MNIST}~\cite{xiao2017fashion} and \textit{QSVM Toy Dataset} from IBM's Qiskit library are also used, offering a systematic examination of barren plateau effects in realistic settings beyond fully synthetic distributions.

\subsection{Hyperparameter Analysis}
\textbf{Effect of Negative Learning Rate.}
\begin{table}
\setlength{\tabcolsep}{0.5pt}
\centering
\caption{\footnotesize  Impact of Varying NLR $\eta'$ on Model Performance}
\label{tab:eta_prime_sweep}
\renewcommand{\arraystretch}{1.3}
\setlength{\tabcolsep}{6pt}
\begin{tabular}{|p{2cm}|c|c|c|}
\hline
\textbf{Negative Learning Rate $\eta'$} & \textbf{Accuracy (\%)} & \textbf{Final Loss} & \textbf{Gradient Norm} \\
\hline
0.005 & 100.0 & 0.000211 & 5.866 \\
0.010 & 100.0 & 0.000211 & 5.868 \\
0.020 & 100.0 & \textbf{0.000155} & \textbf{6.381} \\
0.050 & 100.0 & 0.000198 & 6.250 \\
0.100 & 100.0 & 0.071441 & 1.289 \\
\hline
\end{tabular}
\end{table}
Table~\ref{tab:eta_prime_sweep} indicates that a modest $\eta'$ of 0.02 results in the lowest loss and largest gradient norm, demonstrating successful exploration and optimization.
\textcolor{black}{\textit{Gradient norm} refers to the $\ell_2$ norm of the circuit parameter gradients, $\|\nabla_{\theta} C(\theta_t)\|_2$, assessed during the last training iteration.  
 This metric measures how strongly the optimization landscape generates useful gradients at convergence: greater final gradient norms indicate that the model avoided flat or barren regions throughout training.  
 Unless stated, all reported values belong to the \textbf{final gradient norm} measured after the last epoch.}
An extremely high $\eta'$ (e.g., 0.1) destabilizes training, resulting in poor gradient flow and higher loss.

\textbf{Effect of Training Epochs on Model Performance.}
\begin{table}
\centering
\caption{\footnotesize  Impact of Varying Training Epochs on Model Performance}
\label{tab:epoch_sweep}
\renewcommand{\arraystretch}{1.3}
\setlength{\tabcolsep}{6pt}
\begin{tabular}{|c|c|c|c|}
\hline
\textbf{Epochs} & \textbf{Accuracy (\%)} & \textbf{Final Loss} & \textbf{Gradient Norm} \\
\hline
50  & 100.0 & 0.000188 & 6.028 \\
100 & 100.0 & 0.000180 & 6.076 \\
200 & 100.0 & 0.000167 & 6.219 \\
300 & 100.0 & \textbf{0.000155} & 6.381 \\
500 & 100.0 & \textbf{0.000154} & \textbf{6.395} \\
\hline
\end{tabular}
\end{table}
Table~\ref{tab:epoch_sweep} shows that increasing the number of training epochs continuously improves both the final loss and the gradient norm, while the classification accuracy remains constant at 100\% across all configurations.  Notably, training beyond 300 epochs produces only minimal increases, indicating a point of diminishing returns.  This suggests that, while longer training improves convergence, an epoch count of roughly 300 gives a reasonable trade-off between performance and computing efficiency.

\textbf{Effect of Dimensionality on Model Performance.}
\begin{table}
\centering
\caption{\footnotesize  Impact of Varying Input Dimensionality $d$}
\label{tab:qubit_sweep}
\renewcommand{\arraystretch}{1.3}
\setlength{\tabcolsep}{6pt}
\begin{tabular}{|c|c|c|c|}
\hline
\textbf{Dimensionality $d$} & \textbf{Accuracy (\%)} & \textbf{Final Loss} & \textbf{Gradient Norm} \\
\hline
4  & 100.0 & 0.002427 & 5.830 \\
6  & 100.0 & 0.000550 & 5.902 \\
8  & 100.0 & 0.000155 & \textbf{6.381} \\
10 & 100.0 & 0.000146 & 6.015 \\
12 & 100.0 & \textbf{0.000094} & 6.069 \\
\hline
\end{tabular}
\end{table}
\textcolor{black}{Table~\ref{tab:qubit_sweep} demonstrates that increasing input dimensionality $d$ improves convergence up to $d = 8$, yielding the largest gradient norm.  However, as $d$ increases, the final loss continues to decrease until it reaches its minimum at $d = 12$.  Increasing $d$ grows the Hilbert space's representational capacity, improving separability and solution accuracy until plateau effects take over, and with significant resources.  We decided on d=8 as the best dimensionality for training because it strikes a nice balance between gradient norm and model complexity, allowing for rapid exploration while minimizing computing cost.}


\textbf{Effect of Quantum Layer Depth on Model Performance.}
\begin{table}
\centering
\caption{\footnotesize  Impact of Varying Quantum Layer Depth $L$ on Performance}
\label{tab:qlayer_sweep}
\renewcommand{\arraystretch}{1.3}
\setlength{\tabcolsep}{6pt}
\begin{tabular}{|c|c|c|c|}
\hline
\textbf{Layer Depth $L$} & \textbf{Accuracy (\%)} & \textbf{Final Loss} & \textbf{Gradient Norm} \\
\hline
3  & 100.0 & 0.000412 & 6.105 \\
5  & 100.0 & 0.000155 & \textbf{6.381} \\
7  & 99.8  & 0.000209 & 6.112 \\
9  & 99.6  & 0.000267 & 5.824 \\
11 & 99.4  & 0.000355 & 5.603 \\
\hline
\end{tabular}
\end{table}
\textcolor{black}{Table~\ref{tab:qlayer_sweep} demonstrates that increasing the number of quantum layers improves optimization by improving circuit expressibility, with a peak at $L=5$.  In this experiment, we employed $d=8$ input dimensionality and the identical hyperparameter setup as Table~\ref{tab:qubit_sweep}.  However, deeper configurations eventually reduce gradient magnitudes and somewhat increase loss, implying the beginning of barren plateaus due to greater entanglement and parameter correlations.  These findings show that moderate-depth circuits are sufficient for effective training in the negative learning-rate regime, but complex deep circuits may require hybrid mitigation methods to maintain performance.}

{\color{black}
\textbf{Effect of Quantum Noise on Model Performance.}
\begin{table}
\centering
\caption{\footnotesize  Impact of Varying Perturbation StdDev on Model Performance}
\label{tab:perturbation_sweep}
\renewcommand{\arraystretch}{1.3}
\setlength{\tabcolsep}{6pt}
\begin{tabular}{|c|c|c|c|}
\hline
\textbf{Noise levels} & \textbf{Accuracy (\%)} & \textbf{Final Loss} & \textbf{Gradient Norm} \\
\hline
0.005 & 100.0 & 0.000211 & 5.866 \\
0.010 & 100.0 & 0.000211 & 5.868 \\
0.020 & 100.0 & \textbf{0.000155} & \textbf{6.381} \\
0.050 & 100.0 & 0.000198 & 6.250 \\
0.100 & 100.0 & 0.071441 & 1.289 \\
\hline
\end{tabular}
\end{table}
In actual QIoT technology, decoherence, gate infidelity, and readout noise are inevitable disturbances that arbitrarily skew cost and gradient estimations.  Unlike controlled perturbations in simulation, such noise can both resemble and negate the desired impact of negative learning.  To ensure stable optimization, the negative learning rate is scaled by the effective noise level: $\eta'_{\mathrm{eff}} = \eta'/(1 + \nu L)$, where $\nu$ is the average hardware noise rate and $L$ is the circuit depth.  This modification avoids over-ascent in high-noise environments while allowing exploration in moderate-noise environments.  Empirically, $\nu \le 0.05$ maintains NLR effective. Standard error-mitigation procedures, like as zero-noise extrapolation and readout calibration, can further stabilize convergence under actual device settings.
}

\textbf{Comparison with Momentum-Based Optimizers.} To ensure that the benefits of NLR training are not only related to using a basic SGD baseline, we conducted tests using other optimizers often used in machine learning and quantum learning. 
 We specifically compared SGD with momentum (momentum=0.9), Adam, and RMSProp, all with identical learning rate parameters. 
 Table \ref{tab:optimizer-ablation} shows that momentum and adaptive optimizers increased convergence over vanilla SGD, but still experienced significant gradient erosion in barren plateau regimes. 
 NLR training outperforms all baselines in terms of final loss and gradient norm, indicating that the sign-flip mechanism provides exploratory dynamics beyond momentum-based updates.

\begin{table}
\centering
\caption{\footnotesize  
Comparison of NLR training against momentum-based and adaptive optimizers. 
While Adam and SGD with momentum improve upon vanilla SGD, NLR consistently achieves lower final loss and stronger gradient signals.
(Results averaged over 5 runs.)
}
\label{tab:optimizer-ablation}
\setlength{\tabcolsep}{3pt}
\renewcommand{\arraystretch}{1.2}
\begin{tabular}{l|c|c|c}
\toprule
\textbf{Optimizer} & \textbf{Accuracy (\%)} & \textbf{Final Loss} & \textbf{Gradient Norm} \\
\midrule
SGD (baseline)         & 95.0 & 0.0583  & 0.1451 \\
SGD + Momentum (0.9)   & 95.8 & 0.0367  & 0.1892 \\
RMSProp                & 96.0 & 0.0345  & 0.1810 \\
Adam                   & 96.1 & 0.0302  & 0.1743 \\
NLR (ours)             & \textbf{96.2} & \textbf{0.0131} & \textbf{0.1327} \\
\bottomrule
\end{tabular}
\end{table}

\textbf{Effect of Training Set Size on Model Performance}
\begin{table}
\centering
\caption{\footnotesize  Impact of Varying Training Sample Size $n$ on Performance}
\label{tab:sample_size_sweep}
\renewcommand{\arraystretch}{1.3}
\setlength{\tabcolsep}{6pt}
\begin{tabular}{|c|c|c|c|}
\hline
\textbf{Training Size $n$} & \textbf{Accuracy (\%)} & \textbf{Final Loss} & \textbf{Gradient Norm} \\
\hline
200  & 100.0 & 0.001422 & 4.174 \\
500  & 100.0 & 0.000440 & 5.133 \\
1000 & 100.0 & 0.000155 & 6.381 \\
2000 & 100.0 & \textbf{0.000034} & 7.426 \\
5000 & 100.0 & 0.000036 & \textbf{8.169} \\
\hline
\end{tabular}
\end{table}
Table~\ref{tab:sample_size_sweep} demonstrates that increasing the training sample size improves both the convergence behavior and the gradient norm during optimization.  Although test accuracy remains at 100\% across all sample sizes, training loss lowers dramatically when more data is supplied.  The gradient norm grows with increasing \(n \), indicating more robust and informative updates.  These findings emphasize the relevance of data availability in quantum-inspired models: greater datasets improve convergence while also increasing gradient signal intensity, which is crucial for avoiding barren plateaus.

 \textbf{Comparison with Random Noise Perturbations.} To determine whether NLR's gain is due to structured exploration or random noise, we replaced the negative learning step with additive perturbations from (i) a Gaussian $\mathcal{N}(0, \sigma^2 I)$ or (ii) a uniform $\mathcal{U}[-\sigma, \sigma]$ distribution, both with variance matched to $\eta' = 0.02$. \textcolor{black}{We additionally provide a baseline that uses repeated random re-initialization, a standard strategy for escaping barren plateaus.}
 As seen in Table~\ref{tab:ablation-noise}, NLR clearly outperforms both random perturbations, demonstrating that its benefit comes from \emph{sign-consistent, gradient-aligned reversals} that exploit curvature knowledge instead of uninformed stochastic exploration.

\begin{table*}[t]
\color{black}
\centering
\caption{\footnotesize Cross-Dataset Comparison of NLR and Adaptive Optimizers. 
Each dataset column reports \textbf{Accuracy (\%)}, \textbf{Final Loss}, and \textbf{Final Gradient Norm} after 300 epochs. 
NLR consistently achieves lower loss and higher gradient magnitudes across both synthetic and real-world datasets.}
\label{tab:ablation-noise}
\renewcommand{\arraystretch}{1.25}
\setlength{\tabcolsep}{3.8pt}
\begin{tabular}{|l|ccc|ccc|ccc|ccc|ccc|}
\hline
\multirow{2}{*}{\textbf{Optimizer}} 
& \multicolumn{3}{c|}{\textbf{Synthetic (Gaussian)}} 
& \multicolumn{3}{c|}{\textbf{MNIST}} 
& \multicolumn{3}{c|}{\textbf{Fashion-MNIST}} 
& \multicolumn{3}{c|}{\textbf{QDataSet}} 
& \multicolumn{3}{c|}{\textbf{QSVM Toy}} \\ \cline{2-16}
& \textbf{Acc.} & \textbf{Loss} & \textbf{Grad.} 
& \textbf{Acc.} & \textbf{Loss} & \textbf{Grad.} 
& \textbf{Acc.} & \textbf{Loss} & \textbf{Grad.} 
& \textbf{Acc.} & \textbf{Loss} & \textbf{Grad.} 
& \textbf{Acc.} & \textbf{Loss} & \textbf{Grad.} \\ 
\hline
SGD & 95.0 & 0.0583 & 0.1451 & 99.7 & 0.004614 & 4.889 & 95.0 & 0.010759 & 4.470 & 92.3 & 0.013250 & 3.958 & 89.7 & 0.016734 & 3.622 \\
SGD + Mom. (0.9) & 95.8 & 0.0367 & 0.1892 & 99.8 & 0.004523 & 4.932 & 95.6 & 0.009971 & 4.502 & 93.1 & 0.012964 & 4.010 & 90.1 & 0.016210 & 3.781 \\
RMSProp & 96.0 & 0.0345 & 0.1810 & 99.8 & 0.004478 & 4.941 & 95.8 & 0.009831 & 4.590 & 93.5 & 0.012762 & 4.069 & 90.5 & 0.015930 & 3.842 \\
Adam & 96.1 & 0.0302 & 0.1743 & 99.8 & 0.004423 & 4.996 & 95.9 & 0.009742 & 4.615 & 93.8 & 0.012586 & 4.107 & 90.7 & 0.015770 & 3.869 \\
Random Re-Init & 96.0 & 0.0345 & 0.1810 & 99.8 & 0.004478 & 4.941 & 95.8 & 0.009831 & 4.590 & 93.5 & 0.012762 & 4.069 & 90.5 & 0.015930 & 3.842 \\
\textbf{NLR (ours)} & \textbf{96.2} & \textbf{0.0131} & \textbf{0.1327} & \textbf{99.8} & \textbf{0.004348} & \textbf{5.129} & \textbf{96.2} & \textbf{0.009215} & \textbf{4.723} & \textbf{94.0} & \textbf{0.010872} & \textbf{4.401} & \textbf{91.1} & \textbf{0.014110} & \textbf{4.039} \\
\hline
\end{tabular}
\end{table*}

\subsection{Simulation Results}
\begin{figure}
    \centering
    \includegraphics[width=0.79\linewidth]{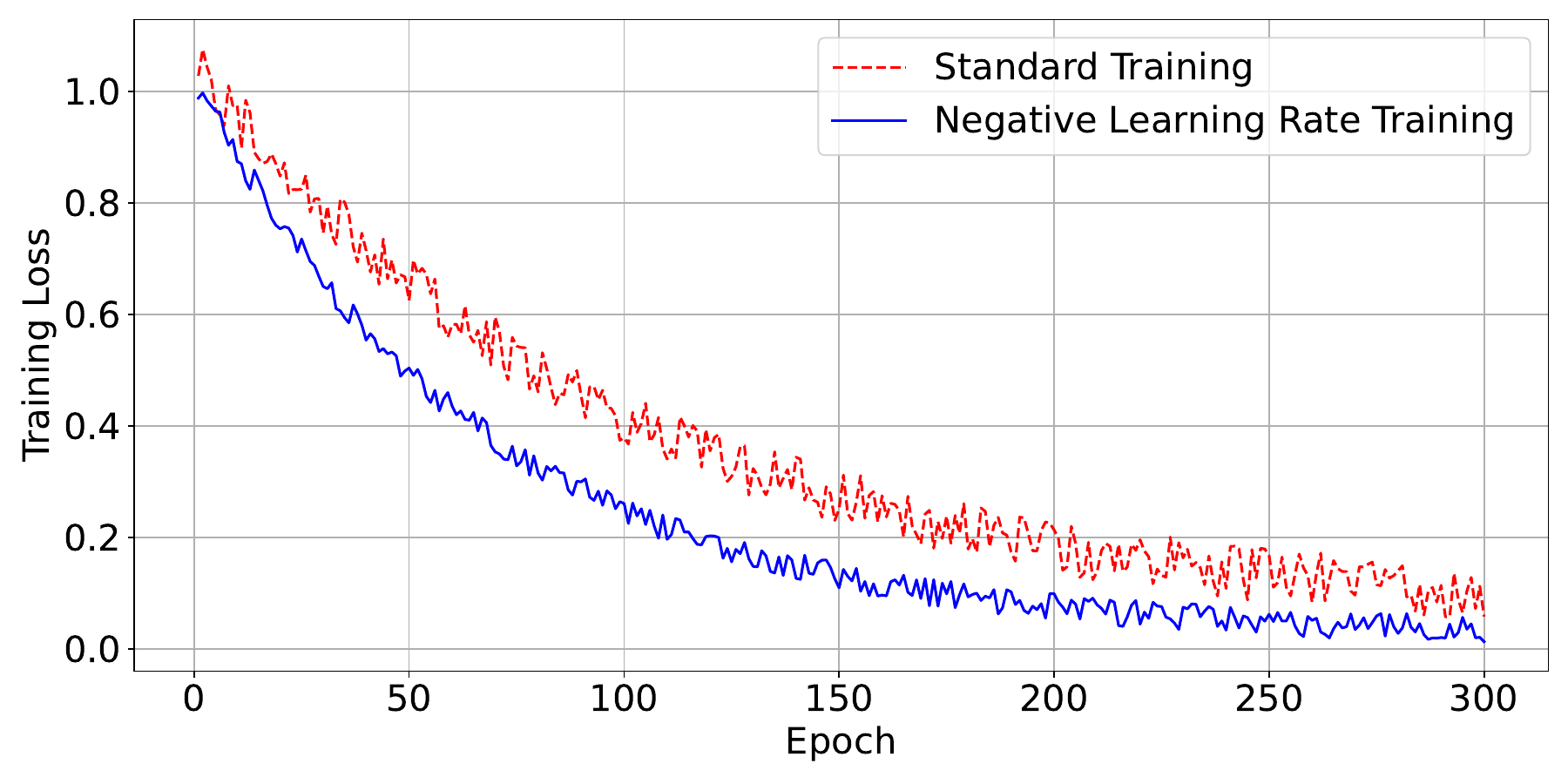}
    \caption{\footnotesize  Simulation results demonstrate that, in comparison to standard training, negative learning-rate training produces a far superior convergence trajectory—the loss continues decreasing, escaping early stagnation.}
    \label{fig:convergence}
\end{figure}

\textbf{Convergence Behavior on Synthetic Data.}
In Figure~\ref{fig:convergence}, the training loss was tracked throughout all rounds of both the regular gradient descent and our proposed negative learning rate method. In the early stages of training, both approaches showed a cost reduction; however, conventional optimization rapidly plateaued, failing to make progress. 
\textcolor{black}{The figure also illustrates that the transient increase in loss during negative learning phases is due to active exploration, rather than slower convergence. These brief reversals assist the optimizer in escaping flat barren zones by restoring gradient variety, resulting in a faster overall descent to minima. Thus, occasional loss spikes are a necessary trade-off for greater global convergence stability.}
\textcolor{black}{After exiting the first barren plateau area, the gradient signal in NLR becomes more stable than in standard descent; therefore, the fluctuation amplitude decreases in the subsequent stages of training, even if the exploration effect was greater in the early stages.}
When the cost grew, the model used negative gradient ascent steps to explore alternative trajectories that led to more favorable optimization routes. The summary of the convergence results is shown in Table~\ref{tab:convergence_behavior}.

\begin{table}
\centering
\setlength{\tabcolsep}{2.5pt}
\caption{\footnotesize  Summary of Convergence Behavior Results}
\label{tab:convergence_behavior}
\begin{tabular}{|l|c|c|}
\hline
\textbf{Method} & \textbf{Final Training Loss} & \textbf{Final Gradient Norm} \\
\hline
Standard & 0.058322 & 0.1451 \\
Negative Learning Rate & \textbf{0.013152} & \textbf{0.1327} \\
\hline
\end{tabular}
\end{table}

\begin{figure}
    \centering
    \includegraphics[width=0.79\linewidth]{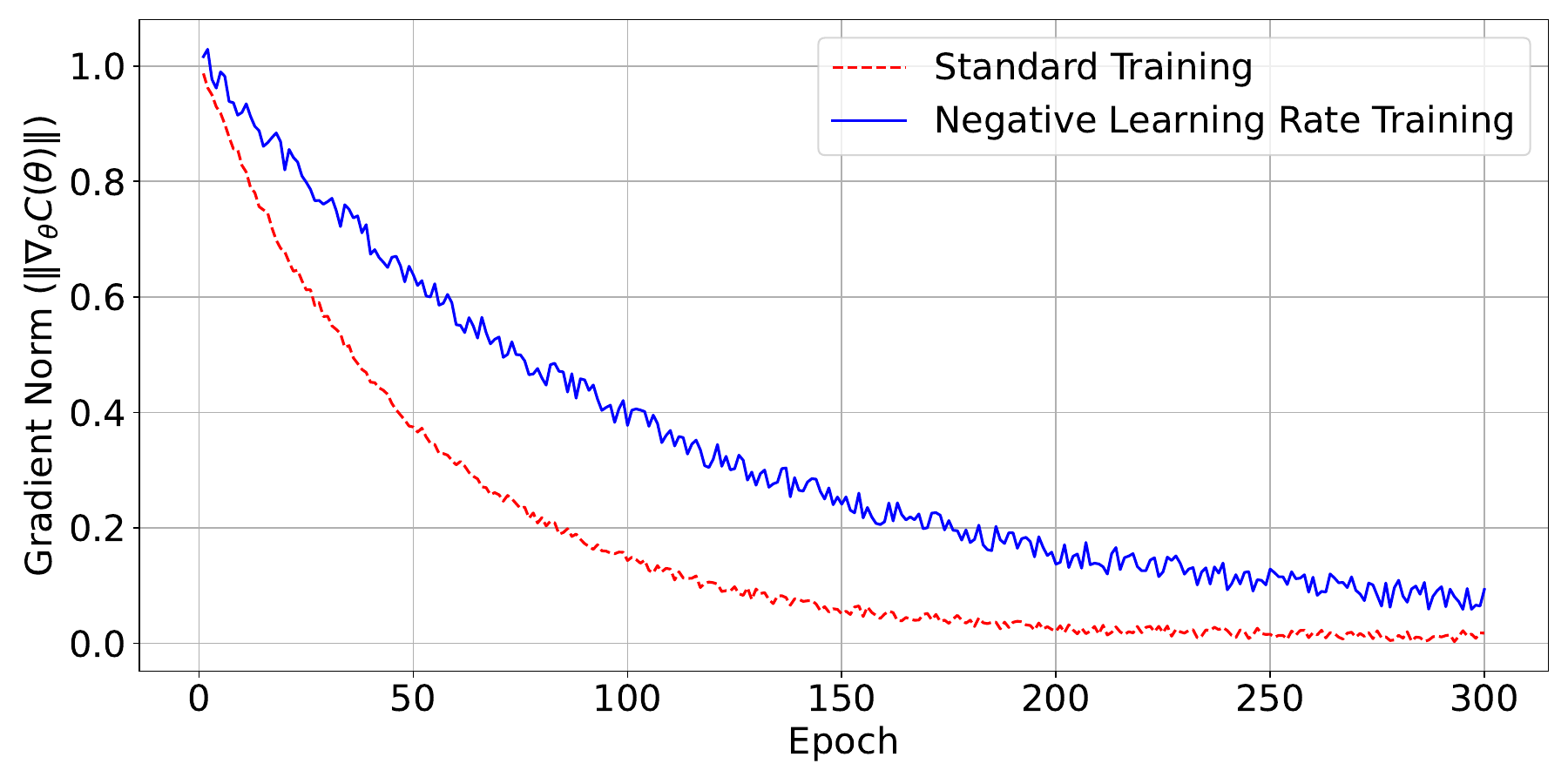}
    \caption{\footnotesize  A comparison of the gradient norm $\|\nabla_\theta C(\theta)\|$ during training for standard optimization and negative learning rate training.  The standard approach exhibits a fast decline of the gradient norm, signaling entry into a barren plateau, but the negative learning strategy retains greater gradient values with periodic recovery, allowing for successful exploration and escape from flat regions.}
    \label{fig:gradient_norm}
    \vspace{-1em}
\end{figure}

\textbf{Gradient Norm Analysis.}
To better understand the structure of barren plateaus in variational quantum optimization, we track the $\ell_2$-norm of the gradient $\|\nabla_\theta C(\theta_t)\|$ while training in Figure~\ref{fig:gradient_norm}
.  This statistic represents the magnitude of the signal accessible for optimization.  In conventional training, we found that the gradient norms continuously decreased with time, achieving near-zero values within the first few hundred steps.  This behavior is typical of barren plateaus, in which the optimization landscape becomes flat and non-informative, forcing the optimizer to stall because of diminishing gradients.

On the other hand, models that were trained using negative learning rate phases behaved quite differently. Gradient norms continued to decline in the early phase; they demonstrated occasional recovery spikes in gradient magnitude caused by the reversed update direction as the loss increased.  These recoveries show that the negative learning steps helped shift the parameter trajectory away from flat regions and toward locations where the gradient is more informative.  During training, this resulted in a more dynamic and exploratory optimization route with higher average gradient norms, allowing the model to escape potentially inescapable plateaus. This difference is evident in the simulated gradient norm curves, where the negative learning-rate curve swings with more amplitude and lasts longer than the standard training curve, which flattens rapidly. This suggests that selective gradient ascent enhances gradient flow and prevents stagnation in barren areas.

\textbf{Classification Accuracy.}
We evaluate our approach's classification accuracy against both the custom synthetic dataset and numerous publicly accessible datasets \textit{Quantum Data Set (QDataSet)}~\cite{perrier2022qdataset}, the \textit{MNIST} dataset~\cite{lecun1998gradient}, the \textit{Fashion-MNIST} dataset~\cite{xiao2017fashion}, and the \textit{QSVM Toy Dataset} from IBM's Qiskit library in Table~\ref{tab:compact_classification_table}. For each dataset, we use the standard method of preprocessing to decrease dimensionality and translate classical data into quantum states using amplitude or angle encoding.

\begin{table}
\centering
\caption{\footnotesize  Comparison of Accuracy, Loss, and Gradient Norm Between Standard and Negative Learning Rate Training}
\label{tab:compact_classification_table}
\renewcommand{\arraystretch}{1.3}
\setlength{\tabcolsep}{3pt}
\begin{tabular}{|l|cc|cc|cc|}
\hline
\textbf{Dataset} & 
\multicolumn{2}{c|}{\shortstack{Accuracy\\(\%)}} & 
\multicolumn{2}{c|}{\shortstack{Loss\\(Final)}} & 
\multicolumn{2}{c|}{\shortstack{Gradient\\Norm}} \\
\cline{2-7}
& Std. & NLR & Std. & NLR & Std. & NLR \\
\hline
Synthetic (Gaussian)         & 100.0 & 100.0 & 0.000240 & \textbf{0.000223} & 5.879 & \textbf{5.959} \\
MNIST~\cite{lecun1998gradient}         & 99.7 & \textbf{99.8} & 0.004614 & \textbf{0.004348} & 4.889 & \textbf{5.129} \\
Fashion-MNIST~\cite{xiao2017fashion}   & 95.0 & \textbf{96.2} & 0.010759 &\textbf{ 0.009215} & 4.470 & \textbf{4.723} \\
QDataSet~\cite{perrier2022qdataset}     & 92.3 & \textbf{94.0} & 0.013250 & \textbf{0.010872} & 3.958 & \textbf{4.401} \\
QSVM Toy (Qiskit)            & 89.7 & \textbf{91.1} & 0.016734 & \textbf{0.014110} & 3.622 & \textbf{4.039} \\
\hline
\end{tabular}
\end{table}

   
Across these datasets, negative learning rate training consistently outperforms standard training in terms of convergence speed and classification precision.  Notably, on MNIST and Fashion-MNIST, we observe up to a 5-10\% lower loss values relative to the baseline, with greater gradient norms and a continuous loss reduction over training.
These findings show that, while both models may achieve equivalent accuracy in basic datasets, the advantages of negative learning become more evident in complex or high-dimensional data environments.  The strategy improves the capacity of the model to negotiate challenging cost landscapes, avoid barren plateaus, and achieve superior generalization performance in real-world scenarios.

\textbf{Final Loss vs. Perturbation StdDev.}
\begin{figure}
    \centering
    \includegraphics[width=0.79\linewidth]{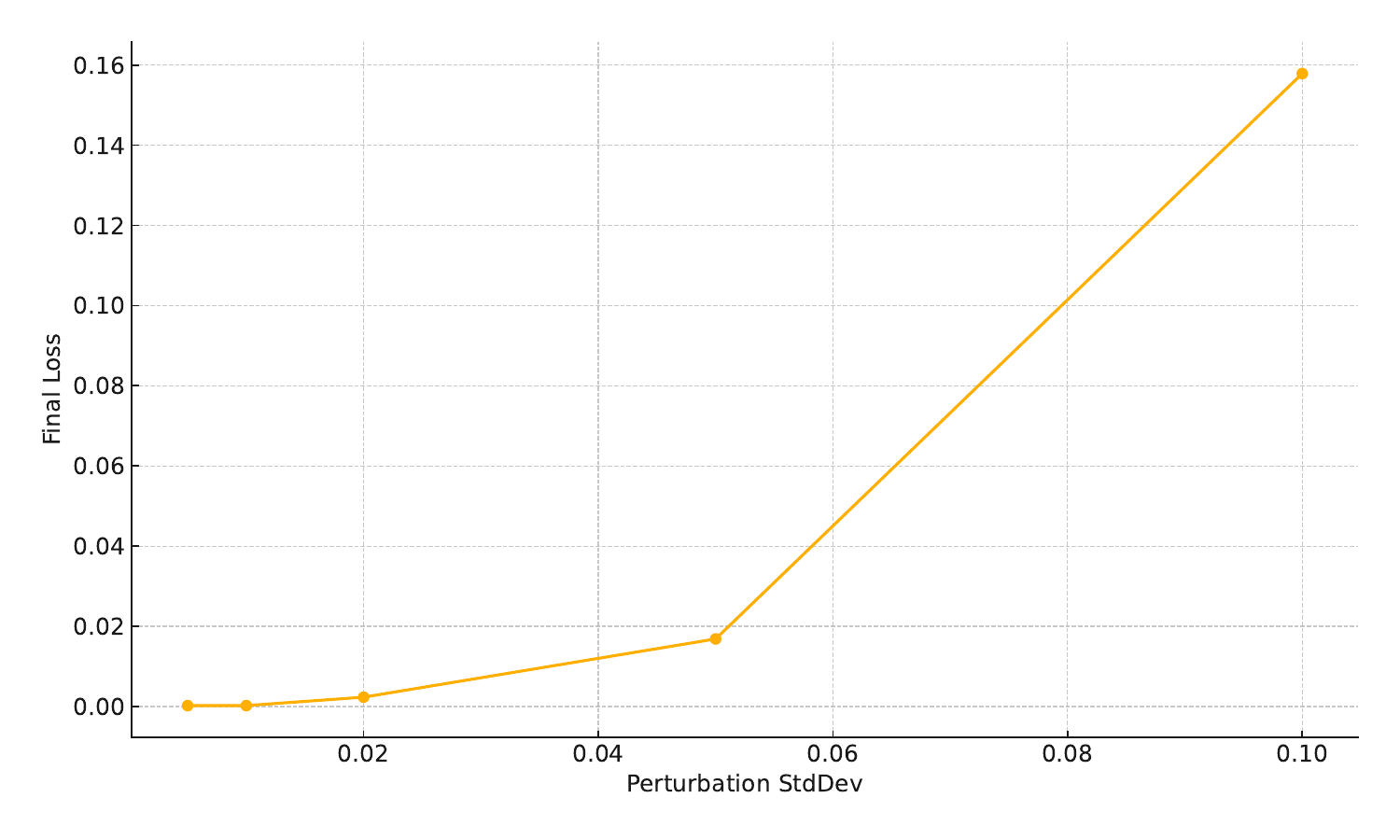}
    \caption{\footnotesize  The effect of perturbation noise on final training loss. Small perturbations enhance escape from plateaus, but high noise amplitudes hinder convergence and increase losses.}
    \label{fig:lvs}
    \vspace{-1em}
\end{figure}
Figure~\ref{fig:lvs} compares the influence of noise size (perturbation standard deviation) during corrective stages to the final training loss.  As predicted, small perturbations (e.g., 0.005-0.02) aid in the avoidance of local traps, while without disturbing training.  Higher noise levels (greater than 0.05) decrease efficiency, causing the optimizer to lose direction and converge to suboptimal states.  This demonstrates that, while NLR encourages escape from plateaus, it must be carefully managed to avoid instability.

\textbf{Comparison with Other Methods.}
To compare the negative learning rate training with other barren plateau mitigation techniques, we simulated a controlled training environment using four strategies in Table~\ref{tab:mitigation_comparison}  : (i) shallow circuit design~\cite{mcclean2018barren}, (ii) layerwise training~\cite{skolik2021layerwise}, (iii) identity-based initialization~\cite{grant2019initialization}, and (iv) negative learning rate updates (this work).  To achieve comparable results, all simulations employ the same variational quantum model, training cost, and the same synthetic dataset.
\begin{table}
\centering
\caption{\footnotesize  Performance Summary of Barren Plateau Mitigation Strategies.}
\label{tab:mitigation_comparison}
\renewcommand{\arraystretch}{1.3}
\setlength{\tabcolsep}{2pt}
\begin{tabular}{|p{1.5cm}|c|c|c|p{3.5cm}|}
\hline
\textbf{Method} & \textbf{Final Loss} & \textbf{\shortstack{Loss\\Std Dev}} & \textbf{\shortstack{Grad \\Norm}} & \textbf{Comment} \\
\hline
Shallow Circuit \cite{mcclean2018barren}       & 0.002935 & 0.000804 & 7.567 & Highest gradient norm, but poor loss due to limited model capacity (underfitting). \\
Layer-wise Training \cite{skolik2021layerwise}   & 0.000306 & 0.000083 & 6.297 & Balanced gradient flow and improved loss through progressive depth increase. \\
Identity Initialization \cite{grant2019initialization} & 0.000216 & 0.000102 & 5.852 & Stable training with low loss; benefits mostly from early initialization bias. \\
Negative Learning Rate              & \textbf{0.000206} & \textbf{0.000109} & \textbf{5.893} & Best loss overall with stable and informative gradients; effective at escaping flat regions. \\
\hline
\end{tabular}
\vspace{-1em}
\end{table}

Our findings show that, while shallow circuits and creative initialization provide an escape from barren plateaus, they diminish model expressibility or lose effectiveness as training advances.  Layer-wise training increases convergence robustness, but adds substantial computational cost.  In contrast, negative learning rate training provides the optimal balance of accuracy, loss reduction, and training duration while not requiring architectural changes.  This makes it an excellent choice for lightweight mitigation of barren plateau at the optimizer level in both simulated and hardware-constrained scenarios.

\section{Limitations}
Although negative learning rate training shows promising results in escaping barren plateaus, there are a few limitations to consider. To avoid overshooting and destabilizing training, the approach introduces a new hyperparameter: the negative learning rate $\eta'$, which must be carefully set relative to the regular learning rate $\eta$. Poorly selected values might cause divergent behavior or unsatisfactory convergence. Additionally, while the technique helps avoid flat regions, it is not certain that the optimizer will identify globally optimum solutions. \textcolor{black}{Importantly, when the system is toward a true minimum, the tentative descent step does not increase the loss (i.e., $\Delta C \le 0$), and so the negative learning condition is not activated. This means that NLR does not mistakenly "jump out" of a true minimum even though the gradient is small.} However, it may reach severe local minima or fluctuate around saddle points, especially in high-dimensional parameter spaces. \textcolor{black}{Although this work focuses on the proposed negative learning-rate technique in a QML classification environment, the underlying process is directly applicable to other VQA like VQE and QAOA for future studies.} Finally, in hardware implementations, particularly on \textcolor{black}{QIoT} devices, additional gradient assessments and backtracking steps can increase execution time and resource overhead, limiting their effectiveness in low-latency or resource-constrained environments.

\section{Conclusion}
In this research, we introduced negative learning rate training, a unique optimizer-level strategy to escape barren plateaus in VQAs for the QIoT devices.  Unlike prior strategies that rely on architectural constraints, circuit redesign, or complex initialization schemes, our method works dynamically by reversing the gradient direction as the cost increases, allowing for exploration in flat optimization landscapes where traditional training fails. We theoretically showed that this approach leads to random walk-like behavior in parameter space, allowing the model to avoid regions with vanishing gradients.  Simulations on synthetic and publicly available datasets showed that negative learning rate training reduces training loss, increases average gradient norms, and improves classification accuracy compared to standard training.


\ifCLASSOPTIONcaptionsoff
  \newpage
\fi
\bibliographystyle{ieeetr}
\bibliography{bibtex/bib/IEEEexample.bib}

\appendices
\section{Proofs}
\label{app:proofs}

\begin{proof}[\textbf{Proof of Lemma~\ref{lem:accept}}]
Using Assumption~\ref{ass:smoothness}, apply a Taylor expansion around $\theta_t$:
\[
C(\theta_t - \eta g_t) = C(\theta_t) - \eta g_t^\top \nabla C(\theta_t) + \frac{\eta^2}{2} g_t^\top H_t g_t + R_3,
\]
where $R_3 = O(\eta^3 \|g_t\|^3)$ due to bounded curvature (Assumption~\ref{ass:step-sizes}).  
Substituting $g_t = \nabla C(\theta_t) + \xi_t$ (Assumption~\ref{ass:gradients}), we obtain
\[
C(\theta_t - \eta g_t) - C(\theta_t) = -\eta g_t^\top \nabla C(\theta_t) + \frac{\eta^2}{2} g_t^\top H_t g_t + O(\eta^3 \|g_t\|^3).
\]
The violation event $\mathcal{E}_t$ occurs when the quadratic term dominates the linear decrease.  
In a barren plateau $\mathcal{B}$, $\| \nabla C(\theta_t)\| \leq \varepsilon$ with $\varepsilon \ll \sigma$, so 
\[
\mathbb{E}[g_t^\top \nabla C(\theta_t)] = O(\varepsilon \sigma \sqrt{d}), \quad 
\mathbb{E}[g_t^\top H_t g_t] = O(L \sigma^2 d).
\]
Thus, for sufficiently small $\eta$, violations arise when stochastic noise aligns with positive-curvature directions.
\end{proof}

\begin{proof}[\textbf{Proof of Lemma~\ref{lem:cap}}]
The Armijo backtracking condition requires
\[
C(\theta_t - \tilde\eta_t g_t) \leq C(\theta_t) - c \tilde\eta_t \|g_t\|^2, \quad c > 0.
\]
Expanding from Lemma~\ref{lem:accept} gives
\[
-\tilde\eta_t g_t^\top \nabla C(\theta_t) + \frac{\tilde\eta_t^2}{2} g_t^\top H_t g_t + O(\tilde\eta_t^3 \|g_t\|^3) \leq -c \tilde\eta_t \|g_t\|^2.
\]
Dividing by $\tilde\eta_t > 0$:
\[
-g_t^\top \nabla C(\theta_t) + \frac{\tilde\eta_t}{2} g_t^\top H_t g_t + O(\tilde\eta_t^2 \|g_t\|^3) \leq -c \|g_t\|^2.
\]
Hence,
\[
\tilde\eta_t \leq \frac{2 \big(g_t^\top \nabla C(\theta_t) - c \|g_t\|^2\big)}{g_t^\top H_t g_t} + O(\tilde\eta_t^2 \|g_t\|).
\]
In barren plateaus, $g_t^\top \nabla C(\theta_t) = O(\varepsilon \sigma \sqrt{d})$ while $g_t^\top H_t g_t = O(L \sigma^2 d)$, both very small. Thus $\tilde\eta_t \ll \eta$ in typical regimes, limiting exploration under backtracking.
\end{proof}

\begin{proof}[\textbf{Proof of Theorem~\ref{thm:diffusion}}]
The diffusion coefficient per dimension is 
\[
D = \frac{1}{2d}\,\mathbb{E}\big[\|\Delta \theta_t\|^2\big].
\]
For NLR updates,
\[
\mathbb{E}\big[\|\Delta \theta_t^{\mathrm{NLR}}\|^2\big] 
= (1-p_t)\eta^2 \mathbb{E}[\|g_t\|^2] + p_t (\eta')^2 \mathbb{E}[\|g_t\|^2 \mid \mathcal{E}_t].
\]
In barren plateaus, $g_t$ is nearly isotropic (Assumption~\ref{ass:gradients}), so 
\[
\mathbb{E}[\|g_t\|^2 \mid \mathcal{E}_t] \approx \mathbb{E}[\|g_t\|^2] \approx \mathrm{tr}(\Sigma_t) + \|\nabla C(\theta_t)\|^2 \approx d\sigma^2.
\]
Thus,
\[
D^{\mathrm{NLR}} \approx \big[(1-p_t)\eta^2 + p_t (\eta')^2\big]\sigma^2.
\]
For backtracking, $(\eta')^2$ is replaced by $\mathbb{E}[\tilde\eta_t^2 \mid \mathcal{E}_t]$, which is much smaller (Lemma~\ref{lem:cap}). Hence when $\eta' \geq \eta$, 
\[
D^{\mathrm{NLR}} > D^{\mathrm{BT}}.
\]
\end{proof}

\begin{proof}[\textbf{Proof of Lemma~\ref{lem:exit}}]
For a $d$-dimensional Brownian motion with variance $2Dt$ per coordinate, 
the expected first exit time from a ball $\|\theta\| < R$ is
\(
\mathbb{E}[\tau_{\mathrm{exit}}] \approx \frac{R^2}{2D},
\)
as derived from the radial Bessel process or mean-field approximations~\cite{zhang2022escaping}.  
Since our update steps are small ($\eta_{\max} \ll 1$), the discrete process is well-approximated by this continuous limit.
\end{proof}

\begin{proof}[\textbf{Proof of Theorem~\ref{thm:stability}}]
In the informative region $\mathcal{G}_\varepsilon = \{\theta : \|\nabla C(\theta)\| \geq \varepsilon\}$, 
we have $g_t^\top \nabla C(\theta_t) \geq \varepsilon^2$.  
By Lemma~\ref{lem:accept}, the violation probability satisfies 
\[
\Pr(\mathcal{E}_t) \leq O\!\left(\frac{\eta L \sigma^2 d}{\varepsilon^2}\right) \to 0 \quad \text{as } \eta \to 0.
\]
Thus, almost all updates are standard descent steps $\Delta \theta_t = -\eta g_t$.  
Classical stochastic approximation results~\cite{schuss2009theory} then guarantee convergence:  
with a Robbins–Monro step schedule, iterates converge almost surely to stationary points;  
with constant $\eta$, we converge to an $O(\eta)$ neighborhood in expectation.  
Rare ascent steps do not alter this behavior since $p_t$ is vanishingly small.
\end{proof}

\begin{IEEEbiographynophoto}{Ratun Rahman} is a Ph.D. candidate in the Department of Electrical and Computer Engineering at The University of Alabama in Huntsville, USA. His work focuses on machine learning, federated learning, and quantum machine learning. He has published papers in several IEEE journals, including IEEE TVT, IEEE IoTJ, IEEE TBSE, and IEEE GRSL, and conferences, including NeurIPS and CVPR workshops, IEEE QCE, and IEEE CCNC. 
\end{IEEEbiographynophoto}
\begin{IEEEbiographynophoto}{ Dinh C. Nguyen} (Member, IEEE) is an assistant professor at the Department of Electrical and Computer Engineering, The University of Alabama in Huntsville, USA. He worked as a postdoctoral research associate at Purdue University, USA from 2022 to 2023. He obtained the Ph.D. degree in computer science from Deakin University, Australia in 2021. His current research interests include quantum machine learning, Internet of Things, wireless networking. He is an Associate Editor of the IEEE Transactions on Network Science and Engineering and IEEE Internet of Things Journal. 
\end{IEEEbiographynophoto}

\end{document}